\documentclass[%
reprint,
superscriptaddress,
 amsmath,amssymb,
 aps,
 pre,
floatfix,
]{revtex4-2}
\usepackage{xcolor}
\usepackage{graphicx}
\usepackage{dcolumn}
\usepackage{bm}

\begin{document}

\preprint{APS/123-QED}
\title{A comparison between D-wave and a classical approximation algorithm and a heuristic for computing the ground state of an Ising spin glass.}
\author{Ran Yaacoby}
\affiliation{Dept.\ \!of Complex Systems, Weizmann Institute of Science, Rehovot 7610001, Israel}%
\author{Nathan Schaar}
\affiliation{Technische Universit\"at Berlin, Chair of Algorithmics and Computational Complexity, Berlin, Germany}%
\author{Leon Kellerhals}
\affiliation{Technische Universit\"at Berlin, Chair of Algorithmics and Computational Complexity, Berlin, Germany}%
\author{Oren Raz}%
\affiliation{Dept.\ \!of Complex Systems, Weizmann Institute of Science, Rehovot 7610001, Israel}%
\author{Danny Hermelin}
\affiliation{Dept.\ \!of Industrial Engineering and Management, Ben-Gurion University of the Negev, Beer-Sheva 8410501, Israel.}%
\author{Rami Pugatch}
\affiliation{Dept.\ \!of Industrial Engineering and Management, Ben-Gurion University of the Negev, Beer-Sheva 8410501, Israel.}%
\affiliation{Quantitative Life Science Section, The Abdus Salam International Center for Theoretical Physics, Strada Costiera 11, 34014, Trieste, Italy}
\date{\today}
\begin{abstract}
Finding the ground state of an Ising-spin glass on general graphs belongs to the class of NP-hard problems, widely believed to have no efficient polynomial-time algorithms for solving them. An approach developed in computer science for dealing with such problems is to devise approximation algorithms that run in polynomial time, and provide solutions with provable guarantees on their quality in terms of the optimal unknown solution. Recently, several algorithms for the Ising-spin glass problem on a graph that provide different approximation guarantees were introduced albeit without implementation.
Also recently, D-wave company constructed a physical realization of an adiabatic quantum computer, and enabled researchers to access it. D-wave is particularly suited for computing an approximation for the ground state of an Ising spin glass on its chimera graph---a graph with bounded degree. In this work, we compare the performance of a recently developed approximation algorithm for solving the Ising spin glass problem on graphs of bounded degree against the D-wave computer. We also compared a heuristic tailored specifically to handle the fixed D-wave chimera graph. D-wave computer was able to find better approximations to all the random instances we studied. Furthermore the convergence times of D-wave were also significantly better. These results indicate the merit of D-wave computer under certain specific instances. More broadly, our method is relevant to other performance comparison studies. We suggest that it is important to compare the performance of quantum computers not only against exact classical algorithms with exponential run-time scaling, but also to approximation algorithms with polynomial run-time scaling and a provable guarantee on performance.
\end{abstract}

\keywords{Adiabatic quantum computation, Ising spin glass, Computational complexity}
\maketitle
\section{\label{sec:Introduction} Introduction}
Finding the ground state of a given quenched disorder realization of the Ising-spin glass Hamiltonian on general graphs belongs to the class of NP-hard problems \cite{Barahona}, widely believed to have no efficient polynomial-time algorithm that  provides the exact ground state. The importance of this problem arises from the ability to map various optimization problems to finding the ground state of a corresponding realization of the Ising spin glass Hamiltonian  \cite{IsingNP}. For example the traveling salesman problem, factoring integers, MAXCUT, graph partition, binary integer linear programming, 3-SAT, Knapsack with integer weights, and graph coloring can all be mapped into finding the ground state of an Ising spin-glass Hamiltonian on some graph \cite{IsingNP}.

Finding the ground state of given Hamiltonian is a natural problem for an adiabatic quantum computer (AQC). An AQC is a physical system with a known controllable Hamiltonian. The initial Hamiltonian has a known and experimentally accessible ground state which is set to be the initial state of the system. After this initialization process, the Hamiltonian's coupling constants $\vec{\lambda}$ are continuously modified. At the end of the process, at $t=t_f$, the system's Hamiltonian is $H_f = H(\vec{\lambda}(t_f))$, which is the Hamiltonian whose ground state is being sought.
According to the adiabatic theorem \cite{RMP}, evolving the coupling constants slowly enough, ensures that at any given time the quantum state of the system $\vec{\psi}(t)$, is the ground state of the instantaneous Hamiltonian $H(\vec{\lambda}(t))$, i.e., that $H(\vec{\lambda(t)}) \vec{\psi}(t) = E_{gs}(\vec{\lambda }(t)) \vec{\psi}(t)$. Thus, upon completing the coupling constants evolution, the state of the system is the desired ground state of the final Hamiltonian $H(\vec{\lambda}_f)=H(\vec{\lambda}(t_f))$, provided that the conditions for adiabaticity are satisfied \cite{RMP}.

Currently, inevitable decoherence and thermalization processes limit the ability of AQCs from always finding the exact ground state of the system. Therefore, current AQCs provide an approximate solution for the ground state. 

In recent years, the D-wave company developed an adiabatic quantum computer which realizes a quantum Ising spin glass Hamiltonian on a bipartite graphs with a certain user-defined range of positive and negative weights on its edges. Finding the classical ground state of an Ising spin-glass on a bipartite graph is known to be NP-hard \cite{Barahona, maxQP}.Therefore, D-wave computer naturally approximates the ground state of an Ising spin-glass problems on a specific bipartite graph (known as the chimera graph, see Fig. 2).  

Since D-wave approximate the ground state of Ising spin-glass problems, it is natural to compare its performance to classical algorithms for approximating the ground state of the same problem. A well-established approach to tackle NP-hard problems in computer science is to devise \emph{approximation algorithms}, which are defined as \emph{algorithms that run in polynomial time, and provide solutions with provable guarantees on their quality in terms of the optimal unknown solution} \cite{CSbook}. This approach is perhaps less familiar among physicists.

Recently, new approximation guarantees for the Ising-spin glass problem on different types of graphs were developed \cite{maxQP}. In particular, for graphs with maximum degree $\Delta$, an algorithm whose run-time scales as $O(n\log n)$, where $n$ is the number of vertices, finds a state with energy that is at most~$2\Delta$ times the energy of the unknown optimal ground state, was suggested in \cite{maxQP}, though not implemented and tested.

In this work, we implemented the approximation algorithm suggested in \cite{maxQP}, which, to the best of our knowledge, is the state of the art in terms of its run-time scaling with the input size. We compared its performance with the performance of the D-wave computer on a predefined set of random instances of the Ising spin-glass ground state problem. In addition, we developed a classical heuristic that exploits the specific structure of the Chimera graph, and improved the approximation at the price of loosing the theoretical guarantee on the accuracy bound. Perhaps surprisingly, both the classical approximation algorithm and the classical heuristic were outperformed by D-wave, which produced better approximations in all the random instances we tested. 

D-wave machine has been used in a variety of studies. Many of the studies embed a problem of interest to the D wave machine and compare it with the state-of-the-art classical algorithm for solving the same problem. The problem with this approach is the overhead that is required in terms of qubits in order to perform the embedding. Other works used the D-wave graph directly, and compared its performance to optimally tuned heuristics, mainly simulated annealing \cite{king2015benchmarking}. Notably, in \cite{PRB15} a classical heuristic was developed that generalizes simulated annealing, yields polynomial run-time and is superior to classical simulated annealing.  

Several researchers studied the question of whether the D wave computer is ``quantun''. In \cite{natphys14} it was demonstrated that D-wave is indeed a ``quantum annealer'', by comparing its performance to an exact quantum Monte-Carlo simulation of D wave system. In contrast, a comparison against classical simulated annealing yielded significantly lower correlations. Later it was shown that D wave performance on random instances can be effectively simulated by a mean-field version of the fully quantum Hamiltonian, which amounts to simulating a classical XZ Hamiltonian coupled to a classical heat bath \cite{shin}. Since the XZ Hamiltonian is obtained from inserting a product form solution to the quantum Hamiltonian used to describe D wave, it was claimed that no entanglement is required for D wave to perform its operation. 

In \cite{PRA92} a special type of bench-marking was proposed. A family of Hamiltonians on the Chimera graph, with a known ground state, and a controllable parameter that changes the location of a second order phase transition was constructed. The critical slowing down of the relaxation time due to the second order phase transition was claimed to cause the problem to be ``hard on average'' \cite{PRA92}.  

In \cite{Troyer}, the authors studied the question of quantum speedup, and argued that the question is quite subtle. Furthermore, they found no evidence for speedup in D wave, although the possibility of speedup was not ruled out.

In this work we do not concern ourselves with the issue of speedup, nor with the question of ``average hardness''. Instead, as mentioned, we look at an approximation algorithm that has a provable performance guarantee, and compare its performance to D wave directly on the chimera graph.  

Our results suggest an intriguing possibility of using the D-wave combined with a classical algorithm to improve the performance of classical approximation algorithms in more general instances of the Ising spin-glass ground state problem. More broadly, we suggest that all quantum algorithms should also be compared with the best known classical approximation algorithm, rather than with classical algorithms that provide the exact solutions or with heuristics without a provable performance bound. 

The structure of this paper is as follows. In section I, we mathematically formulate the Ising spin-glass problem, discuss its computational complexity, describe D-wave's graph structure, and then describe the approximation algorithm and heuristic we use along with the three random ensembles. In the result section we first present the ferromagnetic and anti-ferromagnetic cases where the ground state is known. We then present the main result which is the comparison of D-wave to the classical algorithms across the three ensembles. In the last section we provide an outlook on our results and discuss implications. 

\subsection{\label{sec:Problem} Formulation of the problem of finding the ground-state of the Ising spin glass on a graph}

The Ising spin glass ground state problem is defined as follows. Consider a connected graph $G=(V,E)$ where $V=\{1,2, \ldots,|V|\}$ is the vertex set and $E$ is the edge set of the graph. If $(i,j) \in E$ then vertices $i$ and $j$ are connected in the graph by the edge $(i,j)$. Let $J_{ij}$ be the a real number representing the weight on the edge $(i,j) \in E$ and let $J$ be the symmetric weighted adjacency matrix, and $J_{ij}=0$ when $(i,j)\notin E$. 

We define the Ising spin micro-state vector $\vec{S}$ whose $i$-th component ${S}_i\in \{-1,+1\}$ being the binary spin variable at vertex $i \in V$. The Hamiltonian of the problem is a function that associate a real number with each microstate $\vec{S}$ and is given by 
\begin{eqnarray}
H\{\vec{S}\}=-\sum_{(i,j) \in E} J_{ij} S_i S_j,
\label{eq:one}
\end{eqnarray}
For general weights $J_{ij}$, the problem of finding the ground-state of the above Hamiltonian, i.e. the vector $\vec{S}$ that minimizes the Hamiltonian function $H(\vec S)$, is known as the Ising spin-glass ground state problem. 

Not all instances of the Ising spin glass ground state problem (namely not all instances of $J_{ij}$'s) are difficult, and in fact some instances are trivial. For example, the ferromagnetic case with all the weights being positive, $J_{ij}>0$ for all $(i,j) \in E$, has two trivial solutions: $(\vec{S}_{g.s.})_i = +1$ for all $i \in V$ or alternatively,  $(\vec{S}_{g.s.})_i = -1$ for all $i \in V$.
Additional example for a problem with a trivial solution is a bipartite graph, with all edges negative, i.e., $J_{ij}<0$ for all edges $(i,j) \in E$, i.e., an anti-ferromagnet. In a bipartite graph the vertex set is composed of two disjoint sets of vertices conveniently marked as $L$ and $R$, whose union is the entire vertex set $V$. The edges in a bipartite graph only connects vertices in $L$ to vertices in $R$. Evidently, in this case, the ground state is $S_i=+1$ for $i \in L$ and $S_i=-1$ for $i \in R$ or equivalently, $S_i=-1$ for $i \in L$ and $S_i=+1$ for $i \in R$. 

We also note that the problem of finding the spin configuration with maximum energy in a ferromagnetic setting $(J_{ij}>0)$ is not trivial and belongs to the class of NP-hard problems. This might appear counter intuitive since getting to the highest excited state requires heating, and heating a ferromagnet does not seem to pose a physical challenge. However, heating is not enough, since even in the limit $T \rightarrow \infty$ the system explores all spin configurations with equal probability despite the fact that only one configuration is the maximal energy configuration.

\begin{figure}
\includegraphics[scale=.3]{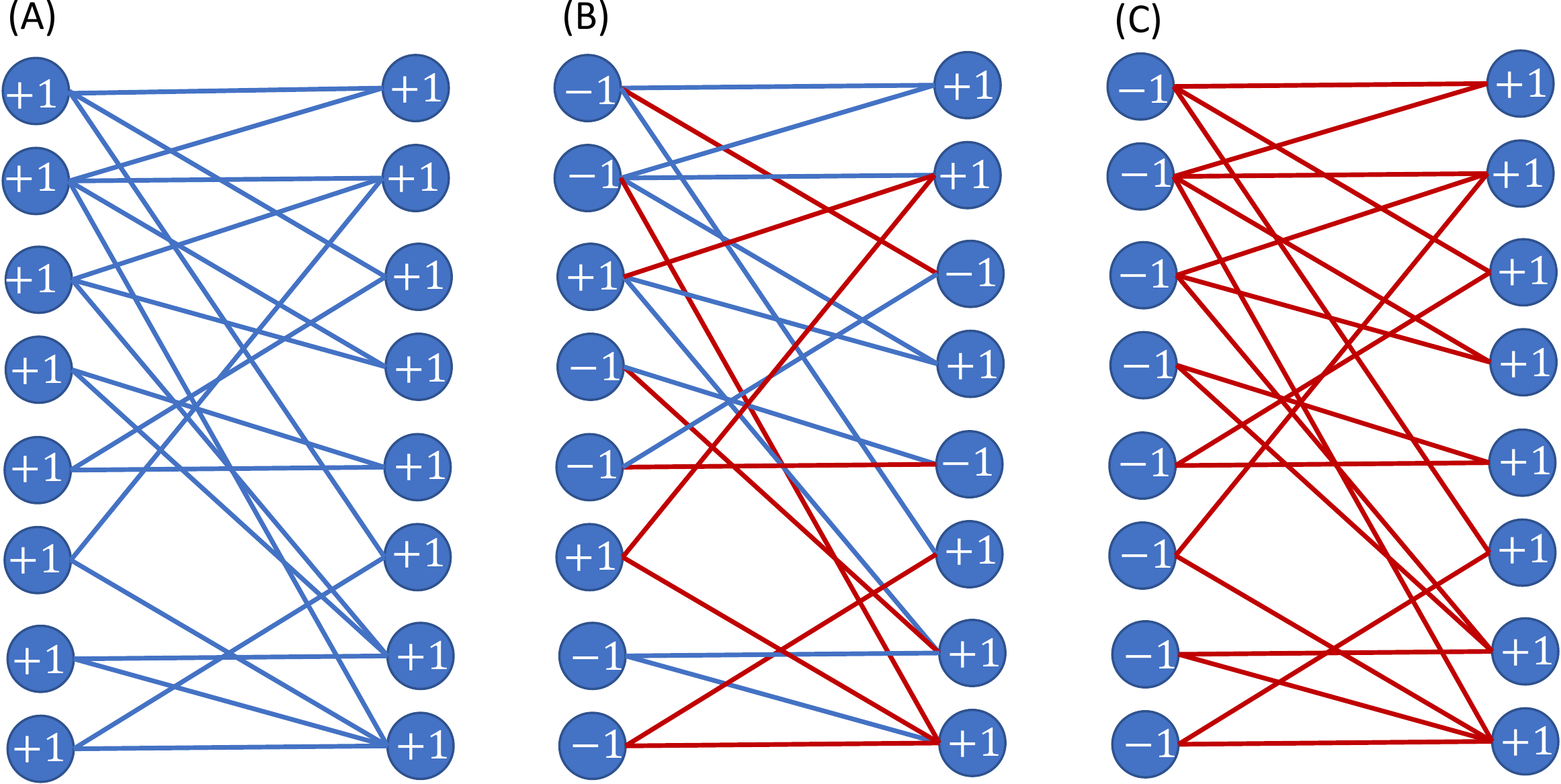}
\caption{\label{fig:bipartite} Panel (A); Finding the ground state energy of Ising spins on a bipartite graph is easy when all the edge weights are positive, i.e., in the ferromagnetic case. Blue edges represent positive weights.  Panel (B); Finding the ground state energy of Ising spins on a bipartite graph with both negative and positive edge weights is NP-hard, i.e., in the spin glass case. Edges colored red represent negative weights. Panel (C); Finding the ground state energy of Ising spins model on a bipartite graph is easy when all the edge weights are positive.}
\end{figure}

Finding the ground-sate of an Ising spin-glass with both positive and negative weights on a general graph is known to belong to the class of NP-hard problems. This is because it includes as a special case the classical max-cut problem, which is the problem of finding a partition of the vertex-set in a given graph into two classes such that the number of edges across the cut is maximized. This problem was shown to be NP-hard by Karp \cite{Karp72} in his celebrated list of 21 initial NP-hard problems.

In the field of combinatorial optimization, approximation algorithms are a popular way for dealing with NP-hard problems. In the context of the Ising spin-glass ground state problem on a graph, an approximation algorithm for a finding the ground state is defined as follows. An approximation guarantee $\alpha \geq 1$ exists, if on any realization of weights $J_{ij}$ of the problem, the approximation algorithm provides a feasible solution $\vec{s}$ with an energy $H(\vec{s}) \leq \alpha E_{g.s.}(J)$. Thus, an approximation algorithm with a guarantee of $1$ always produces the ground state solution. The goal is to devise a \textbf{polynomial-time algorithm} which has the minimal guarantee $\alpha$. 

\subsection{\label{sec:dwave-graph} The D-wave chimera graph}
\begin{figure}
\includegraphics[scale=.45]{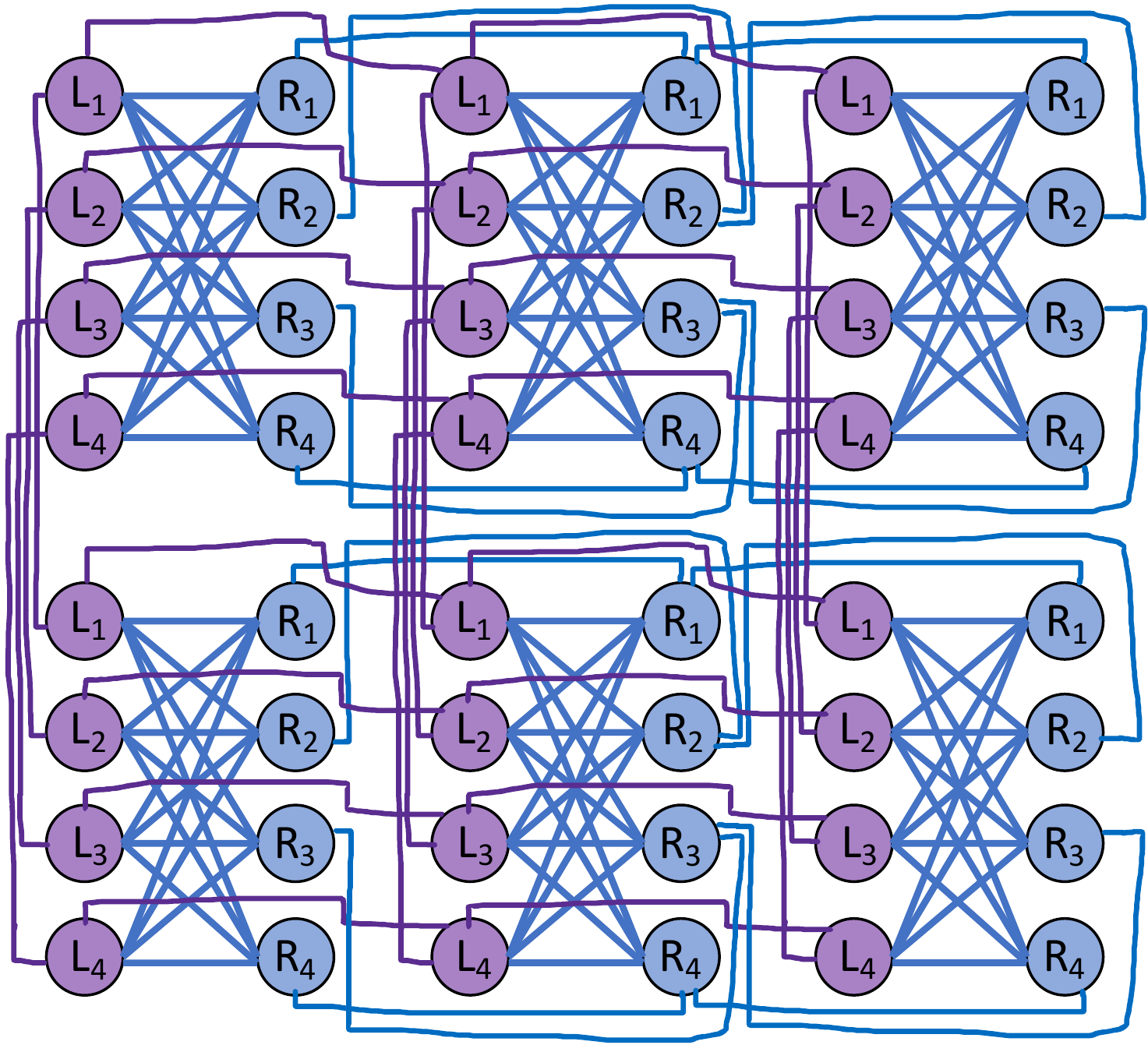}
\caption{\label{fig:chimera} Illustration of a $(2,3)$-chimera graph. The graph contains $2$ rows and $3$ columns of unit-cells---complete-bipartite graphs with $4$ left and $4$ right vertices labeled as $L_k(i,j)$ and $R_k(i,j)$ respectively, with $k \in \{1,2,3,4\}$. Each vertex in a unit cell is connected to a similar vertex in its downward and leftward neighboring unit cell, if it exist. For brevity, we only show one example of such interconnections per direction (left and down) namely, we show that $L_1(1,1)$ connects to $L_1(2,1)$ and we show that $R_4(1,1)$ connects to $R_4(1,2)$. The D-Wave graph is constructed similarly by a sub-graph of a $(4,16)$-chimera graph, with $2041$ vertices, and $5974$ nodes. As mentioned, the maximum degree of a vertex is $6$.}
\end{figure}

Optimally, one would like to compare the performances of classical approximation algorithms to those of an AQC on the most general Ising spin-glass ground state problem. Unfortunately, this is not possible using D wave which is limited to coupling constants that form a ($4,16$)-\emph{chimera} graph.
An ($m,n$)-chimera graph is defined as follow: first consider a complete bipartite graph, with $m$ left vertices and $m$ right vertices. Each left vertex connects to all the right vertices, and vice-versa. The complete bipartite graph is duplicated to form an $n \times n$ grid. Thus, we now have $n^2$ complete bipartite graphs labeled by their grid point $(i,j)$. The left vertices of the complete bipartite graph labeled by the grid point $(i,j)$ are  $L_1(i,j),\ldots,L_m(i,j)$, namely $L_{k}(i,j)$ is the $k$ vertex in the left vertices group in the $(i,j)$ set of vertices in the grid. Similarly, the right vertices are labeled $R_1(i,j),\ldots,R_n(i,j)$ (see Fig.~\ref{fig:chimera}).
Next, the $n^2$ complete bipartite graphs are connected using the following rules: The left vertex $L_k(i,j)$ is connected to $L_{k}(i+1,j)$ for all $i<n$ and for all $j,k$. Similarly, the right vertex $R_{k}(i,j)$ is connected  to $R_{k}(i,j+1)$ for all $j<n$ and for all $i,k$. See Fig.~\ref{fig:chimera} for illustration.

A $(4,16)$ chimera graph thus has $2048$ vertices and~$6016$ edges. D-wave can implement only instances with a sub-graph on $2041$ vertices and~$5974$ edges.
Evidently, the maximum degree of the graph (i.e., the maximum number of vertices connected to any vertex of the graph) is~$6$.

\subsection{\label{sec:approximation} An algorithm with an approximation guarantee}

The approximation algorithm which we compared with D-wave, is an algorithm  first introduced in \cite{maxQP} but implemented here for the first time. It works as follows.
Let $G = (V, E)$ be a graph with edge weights $J_{ij}$, $(i, j) \in E$, and maximum degree $\Delta$. For an edge subset $F \subseteq E$, let $w(F) = \sum_{(i,j) \in F} |J_{ij}|$. We first compute a matching $M \subseteq E$ such that $w(M) \ge w(E)/(2\Delta)$.
We then produce a solution $\vec \psi$ which has energy at most~$-w(M)$, by computing an optimal solution for each pair of vertices matched by $M$, and then carefully extending this solution to the remainder of the graph. Observe that since the ground state has energy at least $-w(E)$, our algorithm has an approximation guarantee of $2\Delta$.

Note that this algorithm has a proven guarantees on the approximation as well as on its run time, however, it does not exploit the specific structure of the chimera graph. We therefore compare to additional algorithm that does exploit this structure, as described next. 

\subsection{\label{sec:heuristic} A heuristic without an approximation guarantee}

The second algorithm we compared with D-wave uses of the structure of the chimera graph described in Sec.~\ref{sec:dwave-graph}.
It first computes an optimal solution $\vec \psi_{i,j}$ for each complete bipartite graph $G_{i,j}$.
Then, it reduces each $G_{i,j}$ to a single vertex.
The edge weights between the $G_{i,j}$ are accumulated so that the resulting graph is a $16\times16$ grid graph $G'$ with new edge weights.
We then compute a solution $\vec \psi'$ for $G'$ using the approximation algorithm above.
Finally, we compute a solution $\vec \psi_{i,j}$, by merging the solutions $\vec \psi_{i,j}$ as follows:
If the solution $\vec \psi'$ assigns the spin $S$ to the grid vertex in row $i$ and column $j$, then we set the spins in $G_{i,j}$ to be $S \cdot \vec \psi_{i,j}$. Note that this algorithm can perform quite badly in the worst-case, especially when the weights of the edges in $G_{i,j}$ are very small compared to the weights of the edges between the $G_{i,j}$'s.

\subsection{\label{sec:Ensembles} Definition of the three random ensembles tested}

To compare the performances of the approximation algorithm and the heuristic described above with D-wave, we need in addition to the graph structure to choose the weights $J_{ij}$ on the graph edges. To this end, we use three random ensembles. The first ensemble is the Gaussian ensemble, where each edge weight $J_{ij}$ on the D-wave graph was drawn from a normal distribution with mean zero and standard deviation equal to $0.1$, $J_{ij} \sim Norm(0,0.1)$, in order to avoid the saturation limit set by D-wave to $1$ in absolute value. The second ensemble is the uniform ensemble, where each edge weight on the D-wave graph $J_{ij}$ was drawn from a uniform distribution ranging from $-1$ to $1$, $J_{ij} \sim [-1,1]$. The third ensemble is the binomial ensemble, where each edge weight on the D-wave graph $J_{ij}$ was equal to $+1$ with probability~$p$, and to~$-1$ with the complementary probability $1-p$.

\section{\label{sec:Results} Results}

\subsection{\label{sec:FAFcases} Comparison between the classical algorithm and D-wave in the ferromagnetic and anti-ferromagnetic cases}
Before comparing the different solutions of the Ising spin-glass ground state problem for difficult instances, where the exact ground state is unknown, we first considered simple instances, as described below.

To each of the three ensembles above, we defined the corresponding ferromagnetic ensemble by using the absolute value of the original weight on each edge, namely $J^{FM}_{ij}=|J_{ij}|$. Clearly, for these ensembles the ground state energy is obtained by setting all the spins to point in the same direction i.e. $S_i=+1$ for all $i \in V$ or $S_i=-1$ for all $i \in V$. Both classical algorithms and D-wave were able to converge to the ground state energy in all tested cases.

Next we tested the anti-ferromagnetic case, i.e. the case with all weights being negative, i.e., $J^{AFM}_{ij}=-|J_{ij}|$. On a general graph, finding the minimal energy is equivalent to finding a weighted max-cut and is thus NP-hard. However, for bipartite graphs, as discussed above, the solution is trivial. Since the vertex set of a bipartite graph is composed of two disjoint groups which we call $L$ and $R$, $V=L \cup R$ and $L\cap R=\{\}$, the optimal solution is to set all spins in $L$ with one sign, and all spins in $R$ with the opposite sign i.e. either $S_i=+1$ for all $i \in L$ and $S_i=-1$ for all $i \in L$, or $S_i=-1$ for all $i \in L$ and $S_i=+1$ for all $i \in L$.

As the D-wave graph is bipartite, we tested its ability to converge to the true ground-state in anti-ferromagnetic instances. We found that in all the tested cases D-wave was able to converge to the true ground-state. Next we tested the ability of the classical algorithms to converge to the true optimum. We found that both the approximation algorithm and the heuristic were able to converge to the true ground state in $95\%$ of the cases.   

\subsection{\label{sec:MainResult} Results of the comparison between D-wave and the classical algorithms on the discrete, uniform and Gaussian weights ensembles}

Next, we consider difficult instances of the problem. For each ensemble, we prepared $1000$ instances with random weights. Each such instance of the Ising spin-glass Hamiltonian was fed as input to the two classical algorithms --- the approximation algorithm with a guaranteed performance bound, and the heuristic algorithm without the performance bound. The same instance was also fed to the D-wave computer. Results were collected and analyzed. In Figures \ref{fig:uniform} to \ref{fig:gaussian}, we depict the results of these comparisons.

In all the ensembles and per each instance the D-wave computer outperformed the two classical algorithms we tested, i.e., D-wave always found a solution with a lower energy compared to the one found by the two classical approximations. The heuristic algorithm without the performance guarantee almost always outperformed the approximation algorithm. Only in rear occasions, and only in the uniform and Gaussian ensembles, the the approximation algorithm outperformed the classical heuristic.

\begin{figure}[t]
\includegraphics[width= 1 \linewidth]{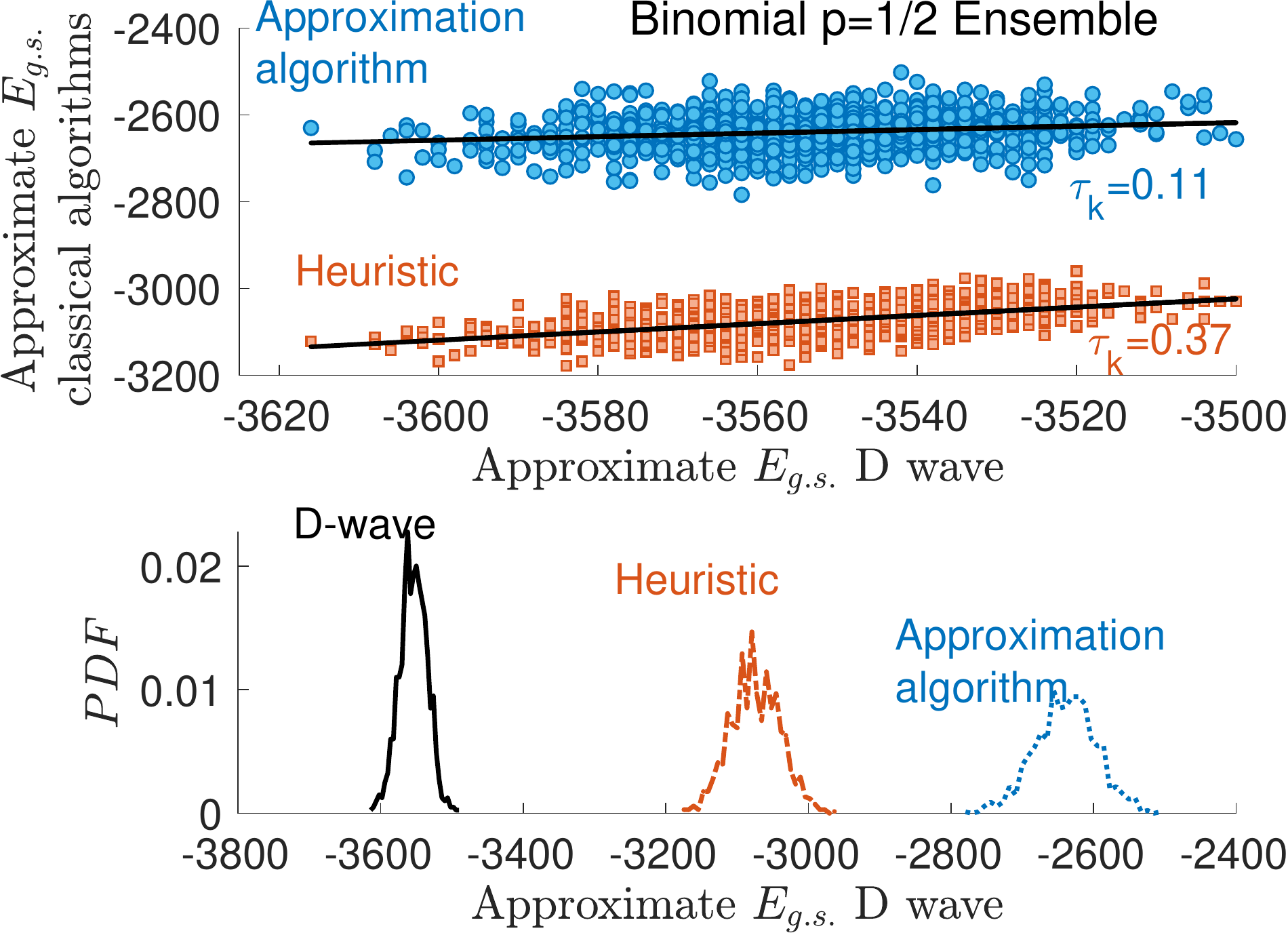}
\caption{\label{fig:discrete} Comparison between D-wave approximated ground state energy and the classical algorithms for the binomial ensemble. The upper graph depicts the energy of each instance obtained by classical algorithms vs D-wave's value. A total of $1000$ random instances of discrete weights on the D-wave graph were tested. The classical approximation algorithm (light blue circles) and the heuristic algorithm (orange squares) are both plotted against the D-wave approximation. The superiority of the D-wave approximation is apparent. For all the instances in this ensemble D-wave found a smaller energy. We calculated Kendal's tau rank correlation and found it to be higher for the heuristic algorithm --- $0.37$ rather than $0.11$ for the approximation algorithm. In the lower graph we present the empirical probability density function of the approximate ground state energy for D-wave (black solid line), Heuristic algorithm (dotted orange line) and the approximation algorithm (dashed light blue line).}
\end{figure}

\begin{figure}[b]
\includegraphics[width= 1 \linewidth]{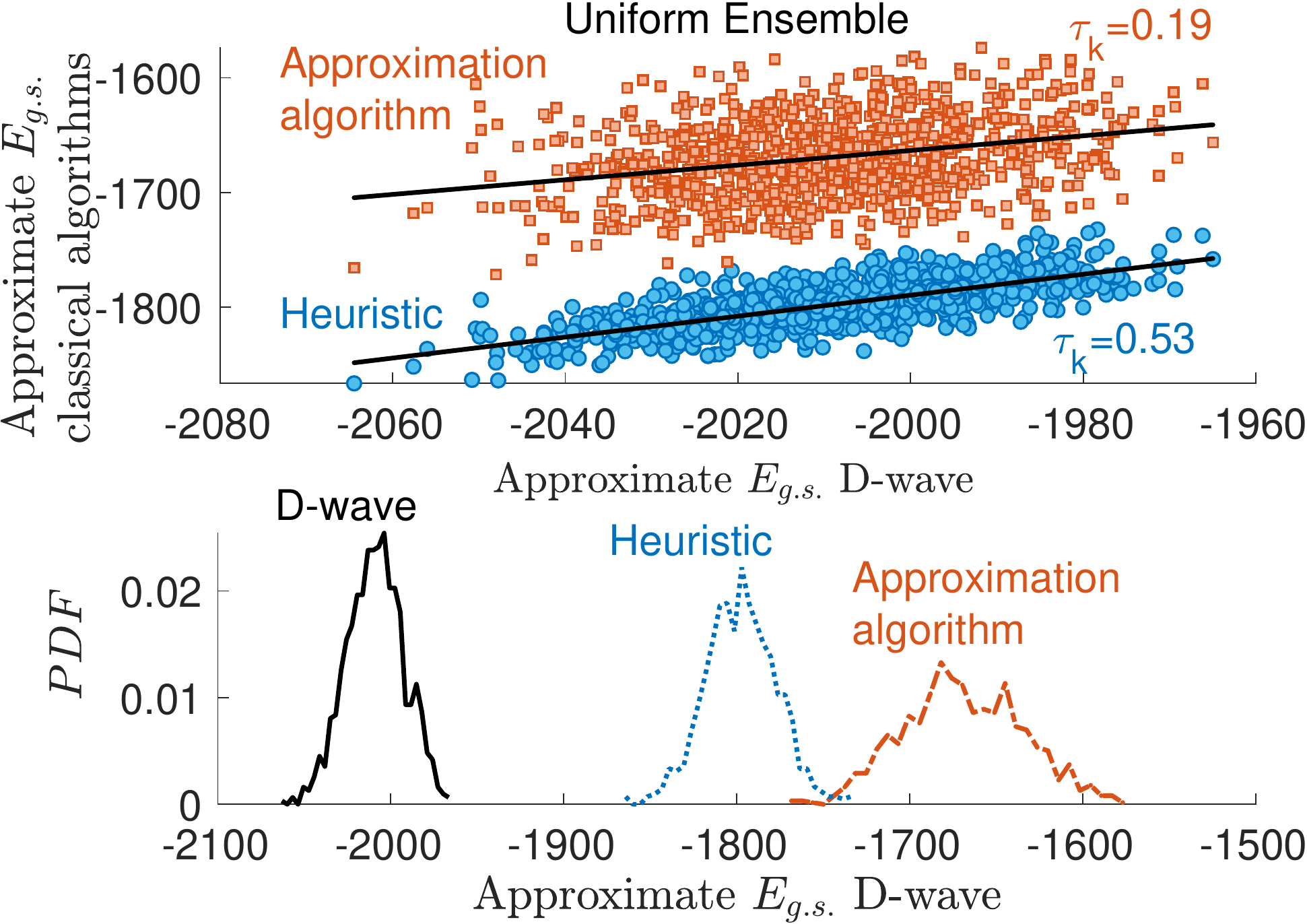}
\caption{\label{fig:uniform} Comparison between D-wave approximated ground state energy and the classical algorithms for the uniform ensemble. The upper graph depicts the energy of each instance obtained by classical algorithms vs D-wave's value. A total of $1000$ random instances of uniform weights on the D-wave graph were tested. The classical approximation algorithm (light blue circles) and the heuristic algorithm (orange squares) are both plotted against the D-wave approximation. The superiority of the D-wave approximation is apparent. For all the instances in this ensemble D-wave found a smaller energy. We calculated Kendal's tau rank correlation and found it to be higher for the heuristic algorithm --- $0.53$ rather than $0.19$ for the approximation algorithm. In the lower graph we present the empirical probability density function of the approximate ground state energy for D-wave (black solid line), Heuristic algorithm (dotted orange line) and the approximation algorithm (dashed light blue line).}
\end{figure}

\begin{figure}[t]
\includegraphics[width= 1 \linewidth]{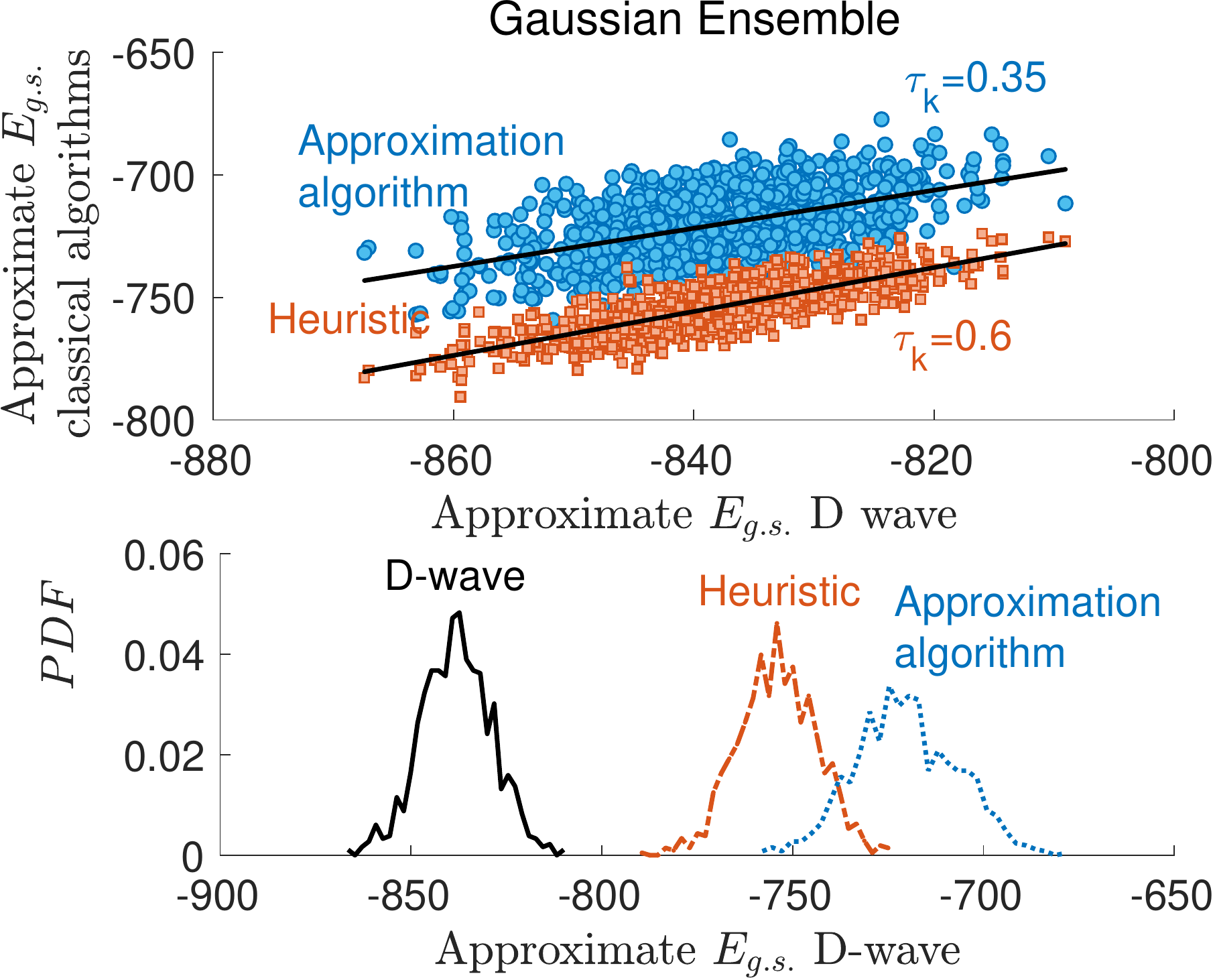}
\caption{\label{fig:gaussian} Comparison between D-wave approximated ground state energy and the classical algorithms for the Gaussian ensemble with mean zero and standard deviation $0.1$ to avoid D-wave's saturation limit set to $1$. The upper graph depicts the energy of each instance obtained by classical algorithms vs D-wave's value. A total of $1000$ random instances of Gaussian weights on the D-wave graph were tested. The classical approximation algorithm (light blue circles) and the heuristic algorithm (orange squares) are both plotted against the D-wave approximation. The superiority of the D-wave approximation is apparent. For all the instances in this ensemble D-wave found a smaller energy. We calculated Kendal's tau rank correlation and found it to be higher for the heuristic algorithm --- $0.6$ rather than $0.3$ for the approximation algorithm. In the lower graph we present the empirical probability density function of the approximate ground state energy for D-wave (black solid line), Heuristic algorithm (dotted orange line) and the approximation algorithm (dashed light blue line).}
\end{figure}

In the binomial ensemble (Fig.~\ref{fig:discrete}), the average improvement of D-wave's ground-state estimate compared to the classical heuristic ranges between $12.5\%$ (minimum improvement) to $19\%$ (maximum improvement) with an average improvement of $15.5 \%$. Furthermore, the average improvement of D-wave ground-state estimate compared to the classical approximation algorithm ranges between $28\%$ to $41\%$ with an average improvement of $35 \%$.

In the uniform ensemble (Fig.~\ref{fig:uniform}), the average improvement of D-wave's ground-state estimate compared to the classical heuristic ranges between $9\%$ to $14.5\%$ with an average improvement of $12 \%$. Furthermore, the average improvement of D-wave ground-state estimate compared to the classical approximation algorithm ranges between $15\%$ to $28\%$ with an average improvement of $20 \%$. 

In the Gaussian ensemble (Fig.~\ref{fig:gaussian}), the average improvement of D-wave's ground-state estimate compared to the classical heuristic ranges between $8\%$ to $14\%$ with an average improvement of $11 \%$. Furthermore, the average improvement of D-wave ground-state estimate compared to the classical approximation algorithm ranges between $11 \%$ to $22\%$ with an average improvement of $16 \%$. 

We conclude that the binomial $p=1/2$ ensemble was more challenging for the classical algorithms compared to D-wave, as D-wave's improvement was more significant in this ensemble. The Gaussian ensemble, on the other hand, was the least challenging for the classical algorithms, as the D-wave improvement was the smallest. 

We also checked the rank correlation between the energies obtained by the classical algorithms to D-wave. If both methods were capable of finding the exact ground state energy, then the rank-correlation between the energies was equal to $1$. The rank correlation is a rough estimate on how well the two algorithms are correlated. A negative correlation would mean very poor performance, zero would mean independent performance, and positive means the two methods agree on the ranking of the energies of the estimated ground states. We found that the approximation algorithm was weakly correlated to D wave, with Kendal's tau rank correlation ranging between $\tau_k=0.11$ for the binomial $p=1/2$ ensemble, to $\tau_k=0.35$ for the Gaussian ensemble. In contrast, Kendal's tau rank correlation between the heuristic algorithm and D wave was ranging between $0.37$ in the binomial $p=1/2$ ensemble, to $0.6$ for the Gaussian ensemble. The improved correlation is consistent with the improved performance of the heuristic algorithm.
 
\section{\label{sec:Discussion} Discussion and outlook}
In this work we compared the performance of the D-wave adiabatic quantum computer against two classical algorithms. The first algorithm was an approximation algorithm that was designed for general sparse graphs which include D-wave's bipartite graph as a special case.


Although the approximation algorithm yields a poor bound, in practice its performance is much better, but nevertheless inferior to the D-wave computer. Our heuristic algorithm which is tailored specifically to the chimera graph, performs better than the approximation algorithm in the majority of cases, but nevertheless was not able to outperform D-wave's computer.

Of course, all our comparisons were made in a specific graph --- the chimera graph. Currently, changing the graph structure is hard as it requires considerable hardware changes. During this research D-wave announced a new generation of chips with a new underlying graph structure called Pegasus. This graph has almost double the number of nodes and more than double the maximum degree as well as a different structure which is incompatible with our heuristic. It would be interesting to test its performance in the future. It will also be of interest to implement the approximation algorithm presented in \cite{AlonNaor06} which has a tighter bound, at the price of also having a run-time scaling which is worse than the one we implemented in this work.

In the future, it will be of interest to find instances for which D-wave performs the worst, in analogy to the approach taken in theoretical computer science. Finding the worst possible weight assignment such that D-wave's approximate ground state energy will have maximal difference from the true ground state, is apparently a hard problem. As a first step towards solving this problem, we suggest to use the method presented in \cite{PRA92} where the ground state energy is known by construction.

Finally, we propose that our method is relevant to all types of performance comparisons between quantum computers and classical algorithms. Rather than just comparing between the performance of a quantum computer and an exact classical algorithm with poor runtime scaling, it is of merit to also compare against a classical approximation algorithm with polynomial runtime scaling and a provable bound on its performance, provided of course it exists \cite{CSbook}.
\begin{acknowledgments}
O. R. is the incumbent of the Shlomo and Michla Tomarin career development chair, and is supported by the Abramson Family Center for Young Scientists and by the Israel Science Foundation, Grant No. 950/19, and the Minerva Stiftung.
R. P. is supported by the Israel Science Foundation Grant No. 776/19.
\end{acknowledgments}

\providecommand{\noopsort}[1]{}\providecommand{\singleletter}[1]{#1}%

\end{document}